\documentclass[sigconf]{acmart}

\usepackage{fancyvrb}
\usepackage{xurl}
\usepackage{booktabs}
\usepackage{url}
\usepackage{amsmath,amsfonts}
\usepackage{algorithmic}
\usepackage{graphicx}
\usepackage{textcomp}
\usepackage{xspace}
\usepackage{subcaption}
\usepackage{tabularx}
\usepackage{framed}
\usepackage{chngpage}
\usepackage[many]{tcolorbox}
\usepackage{pgfplots}
\pgfplotsset{compat=newest}
\usepackage{siunitx}
\usepackage{balance}
\usepackage{enumitem}
\usepackage[leftmargin=0.7cm, rightmargin=0.7cm]{quoting}

\usepackage{hyperref}

\usepackage{tikz}

\usepackage[group-separator={,}, group-minimum-digits=4]{siunitx}

\usepackage{threeparttable}

\newcommand{\rqa}{$RQ_1$}
\newcommand{\rqb}{$RQ_2$}
\newcommand{\rqc}{$RQ_3$}
\newcommand{\rqd}{$RQ_4$}
\newcommand{\rqaa}{Does the frequency of bugs detected through continuous fuzzing differ by language?}
\newcommand{\rqbb}{Do fuzzing bug types and severity differ by language?}
\newcommand{\rqcc}{Does the reproducibility of discovered fuzzing bugs differ by language?}
\newcommand{\rqdd}{Does fuzzing efficiency differ by language?}

\newcommand{\rqA}{\rqa: \rqaa}
\newcommand{\rqB}{\rqb: \rqbb}
\newcommand{\rqC}{\rqc: \rqcc}
\newcommand{\rqD}{\rqd: \rqdd}

\newcommand{\rqaaOne}{Vulnerability Ratio}

\newcommand{\rqbbOne}{Crash Type}
\newcommand{\rqbbTwo}{Severity Level}

\newcommand{\rqccOne}{Patch Coverage}
\newcommand{\rqccTwo}{Time To Detection}

\newcommand{\ToIssueCollectDate}{2025-06-15\xspace}      

\newcommand{\TotalIssueReportOnlyNum}{68495}    
\newcommand{\ClosedIssuesOnlyNum}{63652}        
\newcommand{\RegressedIssuesOnlyNum}{43002}        
\newcommand{\RegressedProjectsOnlyNum}{345}

\newcommand{\TotalIssueReport}{\num{\TotalIssueReportOnlyNum}\xspace}
\newcommand{\ClosedIssues}{\num{\ClosedIssuesOnlyNum}\xspace}
\newcommand{\RegressedIssues}{\num{\RegressedIssuesOnlyNum}\xspace}
\newcommand{\RegressedProjects}{\num{\RegressedProjectsOnlyNum}\xspace}

\newcommand{\TotalFuzzingProjectNum}{1311}
\newcommand{\TargetProjectNum}{563}
\newcommand{\TotalFuzzingProject}{\num{\TotalFuzzingProjectNum}\xspace}
\newcommand{\TargetProject}{\num{\TargetProjectNum}\xspace}

\pgfmathsetmacro{\TargetProjectRateNum}{(\TargetProjectNum/\TotalFuzzingProjectNum)*100}

\newcommand{\TargetProjectRate}{\pgfmathprintnumber[precision=1]{\TargetProjectRateNum}\xspace}

\newcommand{\AllBuildLog}{\num{4660639}\xspace}

\newcommand{\FinalProject}{\num{559}\xspace}
\newcommand{\FinalIssue}{\num{61444}\xspace}
\newcommand{\FinalBuildLog}{\num{999248}\xspace}
\newcommand{\FinalCoverageReport}{\num{669015}\xspace}

\newcommand{\rqOneVulAllRate}{23.48\xspace}

\newcommand{\PatchCoverageProject}{\num{312}\xspace}
\definecolor{darkgreen}{rgb}{0, 0.5, 0} 
\definecolor{whitesmoke}{rgb}{0.99, 0.99, 0.99} 

\newcommand{\red}{\color{black}}
\newcommand{\blue}{\color{black}}

\newcommand{\black}{\color{black}}

\def\Underline{\setbox0\hbox\bgroup\let\\\endUnderline}
\def\endUnderline{\vphantom{y}\egroup\smash{\underline{\box0}}\\}
\def\|{\verb|}

\newcommand{\ie}{\textit{i.e.,}\xspace}
\newcommand{\eg}{\textit{e.g.,}\xspace}

\newcommand{\etal}{\xspace\textit{et al.}\xspace}

\newcounter{findings_no}

\usepackage{listings}
\definecolor{backcolour}{rgb}{0.95,0.95,0.92}
\lstdefinelanguage{diff}{
  morecomment=**[f][\color{red}]{-},         
  morecomment=**[f][\color{darkgreen}]{+},       
  moredelim=**[is][\bfseries]{@@}{@@},
}
\definecolor{backcolour}{rgb}{0.95,0.95,0.92}
\lstdefinelanguage{commit}{ 
  breakindent = 0pt,
  numbers=none,
  backgroundcolor=\color{white},
  frame=single,
  xleftmargin=3.5em,
  numbersep=0em,
  xrightmargin=1.5em,
}

\usepackage[many]{tcolorbox}  
\tcbuselibrary{listings,breakable}
\definecolor{main}{HTML}{D0D3D4}    
\definecolor{sub}{HTML}{D0D3D4}     
\newtcolorbox{dbox}{
    left=1pt,right=1pt,top=2pt,bottom=2pt,
    enhanced, 
    boxrule = 0pt,
    enlarge top by=5pt,
    enlarge bottom by=3pt,
  }
\tcbset{
    sharp corners,
    before skip = 0.1cm,    
    after skip = 0.01cm      
}

\AtBeginDocument{%
  \providecommand\BibTeX{{%
    \normalfont B\kern-0.5em{\scshape i\kern-0.25em b}\kern-0.8em\TeX}}}

\copyrightyear{2026}
\acmYear{2026}
\setcopyright{cc}
\setcctype{by}
\acmConference[MSR ’26]{23rd International Conference on Mining Software Repositories}{April 13--14, 2026}{Rio de Janeiro, Brazil}
\acmBooktitle{23rd International Conference on Mining Software Repositories (MSR ’26), April 13--14, 2026, Rio de Janeiro, Brazil}
\acmPrice{}
\acmDOI{10.1145/3793302.3793358}
\acmISBN{979-8-4007-2474-9/2026/04}

\acmConference[MSR '26]{The 23rd International Mining Software Repositories Conference}{April 13--14,
  2026}{Rio de Janeiro, Brazil}
\acmISBN{978-1-4503-XXXX-X/18/06}

\begin{document}

\title{Does Programming Language Matter? An Empirical Study of Fuzzing Bug Detection}






\author{
  Tatsuya Shirai\textsuperscript{$\dagger$},
  Olivier Nourry\textsuperscript{$\ddagger$},
  Yutaro Kashiwa\textsuperscript{$\dagger$},
  Kenji Fujiwara\textsuperscript{$\S$},
  Hajimu Iida\textsuperscript{$\dagger$}
}

\affiliation{%
  \institution{\textsuperscript{$\dagger$}Nara Institute of Science and Technology, Japan}
  \country{}
}
\affiliation{%
  \institution{\textsuperscript{$\ddagger$}The University of Osaka, Japan}
  \country{}
}
\affiliation{%
  \institution{\textsuperscript{$\S$}Nara Women’s University, Japan}
  \country{}
}

\email{{shirai.tatsuya.sp1, yutaro.kashiwa, iida}@naist.ac.jp}
\email{nourry@ist.osaka-u.ac.jp,kenjif@ics.nara-wu.ac.jp}

\authorsaddresses{%
  Authors' addresses: 
  T. Shirai, Y. Kashiwa, H. Iida, Nara Institute of Science and Technology, Japan; 
  O. Nourry, The University of Osaka, Japan; 
  K. Fujiwara, Nara Women's University, Japan.
}

\renewcommand{\shortauthors}{Shirai, et al.}

\begin{abstract}
Fuzzing has become a popular technique for automatically detecting vulnerabilities and bugs by generating unexpected inputs. In recent years, the fuzzing process has been integrated into continuous integration workflows \blue (\ie continuous fuzzing), \black enabling short and frequent testing cycles. Despite its widespread adoption, prior research has not examined whether the effectiveness of continuous fuzzing varies across programming languages.

This study conducts a large-scale cross-language analysis to examine how fuzzing bug characteristics and detection efficiency differ among languages. We analyze \FinalIssue fuzzing bugs and \FinalBuildLog builds from \FinalProject OSS-Fuzz projects categorized by primary language. Our findings reveal that (i) C++ and Rust exhibit higher fuzzing bug detection frequencies, (ii) Rust and Python show low vulnerability ratios but tend to expose more critical vulnerabilities, (iii) crash types vary across languages and unreproducible bugs are more frequent in Go but rare in Rust, and (iv) Python attains higher patch coverage but suffers from longer time-to-detection. These results demonstrate that fuzzing behavior and effectiveness are strongly shaped by language design, providing insights for language-aware fuzzing strategies and tool development.

\end{abstract}



\maketitle

\renewcommand{\smallskip}{\vspace{2pt}}

\section{Introduction}\label{sec:introduction}


Fuzzing is a testing methodology that automatically generates a large number of inputs and feeds them to a target software system to trigger unintended behaviors (\eg crashes or memory overflows) and uncover defects.
The effectiveness of fuzzing has been demonstrated by multiple studies~\cite{DBLP:journals/compsec/CuiCHLDL22,DBLP:conf/sigsoft/BohmeF20}.

To make fuzz testing more accessible in modern software development workflows, continuous fuzzing integrates fuzzing directly into continuous integration (CI) pipelines with frequent, short testing sessions. This approach addresses the traditional challenges of setting up complex fuzzers and managing high computational costs~\cite{DBLP:journals/tosem/NourryKLBLK24}. Rather than running for weeks on a single software version, continuous fuzzing operates daily on actively developed code. 
Klooster\etal\cite{DBLP:conf/icse/KloosterTBHB23} conducted a study \blue on the effectiveness and scalability of fuzzing within \black CI pipelines and demonstrated that even 15-minute sessions can uncover critical bugs.


In recent years, several studies have investigated continuous fuzzing practices and software defects detected through continuous fuzzing (hereafter referred to as fuzzing bugs). Ding\etal\cite{DBLP:conf/msr/DingG21} analyzed the characteristics of fuzzing bugs, finding that many of them cause the software to become unavailable to users, \eg by causing crashes or resource exhaustion, and that non-reproducible bugs are less likely to be fixed. Keller\etal\cite{DBLP:conf/msr/KellerMM23} also conducted a lifecycle analysis of fuzzing bugs, revealing that while the median lifespan of fuzzing bugs was 324 days, they were addressed within a median of only 2 days after detection.

While these prior studies have provided interesting new insights on continuous fuzzing, their analyses do not distinguish between programming languages or focus on specific languages. Programming languages vary significantly in their design principles and safety guarantees. For example, Rust emphasizes memory safety through compile-time checks, while others like C/C++ provide low-level memory control at the cost of safety. Given these fundamental differences, it is still not clear whether prior findings, derived from language-agnostic analyses, can be directly generalized across multiple languages. In this context, we thus pose the central question:

\vspace{0.6mm}
\begin{quoting}
    \textbf{\textit{``Do fuzzing sessions/bugs manifest differently across languages?''}}
\end{quoting}
\vspace{0.6mm}

In this study, we perform a cross-language analysis to investigate how fuzzing bug characteristics and detection efficiency vary across programming languages. We analyze \FinalIssue fuzzing bugs, and \FinalBuildLog builds from \FinalProject OSS-Fuzz repositories, categorized by primary language. Our contributions are:
\begin{enumerate}[leftmargin=15pt,itemsep=2pt, parsep=0pt, topsep=2pt]
    \item \textbf{Programming Languages show distinct bug detection Distributions:} Although the median bug detection rate is generally low across all languages (around 0.02 issues per build), the frequency distribution differs significantly across languages. Specifically, C++ and Rust exhibit the widest distributions and the highest variability, with long tails extending far beyond the median, while Python shows the lowest median detection rate (0.014) and the narrowest, most concentrated pattern. 
    
    \item \textbf{Language differences shape vulnerability patterns:} The nature of vulnerabilities differed sharply across languages. Specifically, Java projects exhibited a distinct profile dominated by medium-severity (S3) vulnerabilities, contrasting with the predominance of \red critical and high-severity (S1/S2) vulnerabilities \black in C, C++, Go, Python, and Rust. Furthermore, C, C++, and Go exhibit mainly Resource Management bugs, whereas Java, Python, and Rust tend to show more Control Flow bugs.

    \item \textbf{Rust demonstrates superior fuzzing bug reproducibility:} Rust projects demonstrate superior fuzzing bug reproducibility. These projects showed the best reproducibility ratio of fuzzing bugs (median around 98.8\%) and a tight distribution near 100\%. Additionally, while Rust projects detect numerous fuzzing bugs, their vulnerability ratio remains low (below 10\%).

    \item \textbf{Fuzzing efficiency depends on language characteristics:} Fuzzing efficiency, measured by Patch Coverage and Time-to-Detection (TTD), varies significantly across languages. Interestingly, languages with relatively high patch coverage, such as Go and Python, tend to show longer time to detection, while C, C++, Java, and Rust exhibit lower patch coverage but shorter detection times. This suggests that the expected relationship where testing newly modified code (high patch coverage) facilitates early bug detection does not universally hold across all programming languages when using fuzzing.

    \item \textbf{Replication Packages:} To facilitate replication and further studies, we provide the data used in our replication package.\footnote{\url{https://doi.org/10.5281/zenodo.18264483}} 
\end{enumerate}


\section{Background}\label{sec:background}
\subsection{Continuous Fuzzing}

Continuous fuzzing integrates the fuzzing process directly into the CI pipeline, running automated fuzz tests regularly (\eg daily). While traditional fuzzing is typically performed manually (\eg before releasing), continuous fuzzing executes in short, frequent cycles. This approach enables teams to quickly and automatically detect bugs as soon as they enter the codebase, rather than discovering them later during development.

OSS-Fuzz, provided by Google, is the most widely used continuous fuzzing service, currently integrating and fuzzing over 1,000 critical open-source software projects. Since its launch in 2016, OSS-Fuzz has helped identify and fix over 13,000 vulnerabilities and 50,000 bugs across participating projects. It officially supports six languages, including C, C++, Rust, Go, Python, Java, and JavaScript,\footnote{https://github.com/google/oss-fuzz} enabling fuzzing across many critical projects such as OpenSSL, Kubernetes, and TensorFlow.

\begin{figure}[t]
  \centering
  \includegraphics[width=0.99\linewidth]{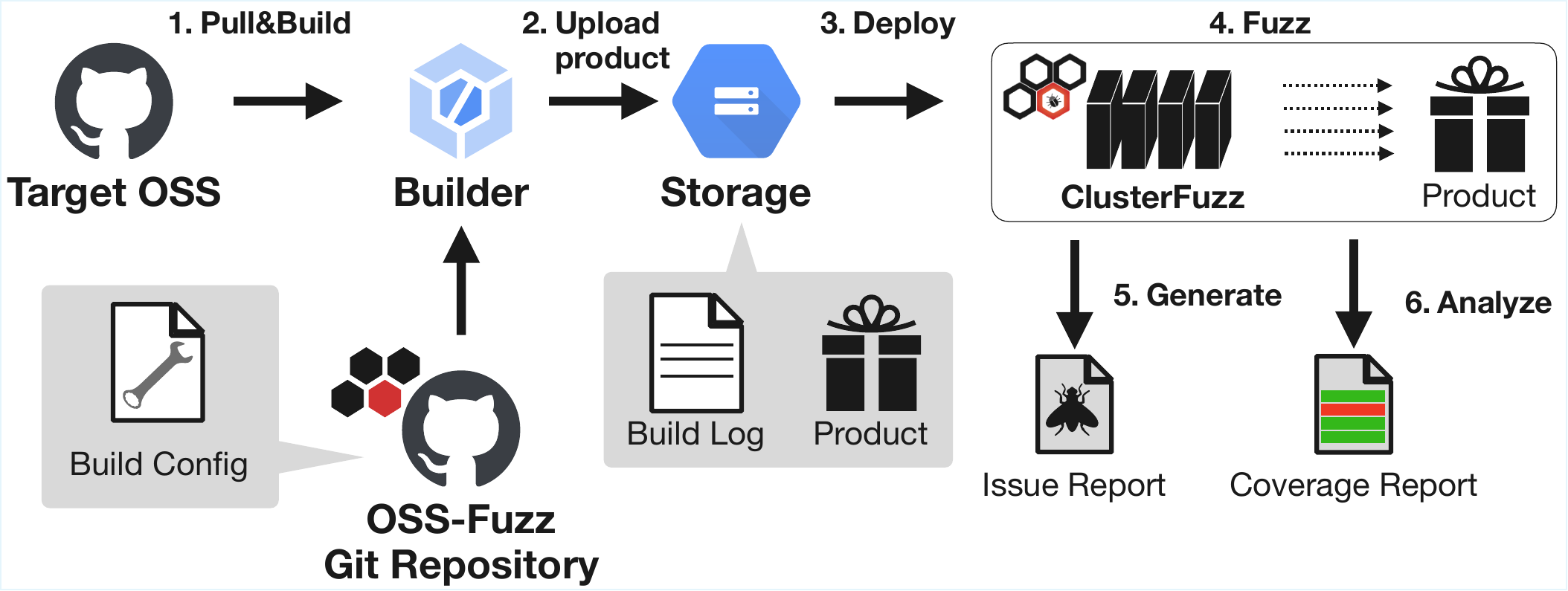}
  \caption{Architecture of OSS-Fuzz}
  \label{fig:OSS-Fuzz}
\end{figure}
\autoref{fig:OSS-Fuzz} illustrates the bug detection process using OSS-Fuzz. To use the service, developers must ensure that their repository is public, must write their own fuzz targets, and have to upload their build configuration files to OSS-Fuzz. Using these build files, OSS-Fuzz will then be able to build the target open-source project and its fuzz targets to carry out the fuzzing process.

After the build completes, a fuzzing build log is generated and uploaded to a Google Cloud Storage (GCS) bucket.\footnote{\url{https://oss-fuzz-build-logs.storage.googleapis.com/log-61e82104-9c63-41eb-83f6-6b0bfcf861e3.txt}} The compiled project files and fuzz targets are also uploaded to cloud storage for use by OSS-Fuzz's fuzzers.

Next, ClusterFuzz, the distributed fuzzing infrastructure underlying OSS-Fuzz, downloads the compiled source files along with the initial fuzzing inputs (\ie the fuzzing corpus) and begins fuzzing. By default, ClusterFuzz uses \texttt{libfuzzer}\footnote{\url{https://llvm.org/docs/LibFuzzer.html}} as its fuzzer, and more than 99\% of projects in OSS-Fuzz use either libfuzzer by itself or in combination with other fuzzers. When ClusterFuzz's fuzzers trigger a bug during fuzzing, an issue report is automatically generated and submitted to the issue tracking system.\footnote{\url{https://issues.oss-fuzz.com/issues?q=status:closed}}

Each issue report includes the project name, the fuzzing engine used during the execution, the crash information, the severity of the bug, the regression range (\ie the range of commits where the bug was introduced), and the fixed range (\ie the range of commits where the bug was fixed). Reports remain private for 90 days from the reporting date or until the bug is fixed by the projects.

Additionally, once the fuzzing process is completed, a daily coverage report is generated which provides various fuzzing coverage information such as the hit frequency, the fuzzer reachability reports, and the aggregated coverage reports to understand the performance of the fuzzer and identify any blockers~\cite{fuzz-inspector}. These coverage reports are also uploaded to the OSS-Fuzz GCS bucket and made available publicly.\footnote{\url{https://storage.googleapis.com/oss-fuzz-coverage/openssl/reports/20250613/linux/report.html}}

\subsection{Related Work}
We categorize related work into three approaches: language-agnostic studies (no differentiation between languages), language-specific studies (analyzing fuzzing practices and bugs for a particular language), and cross-language studies (examining specific bug types across different languages).

\subsubsection{Language-Agnostic Analysis}
Many studies have examined the effectiveness of continuous fuzzing~\cite{DBLP:conf/msr/KellerMM23}, the characteristics of detected bugs~\cite{DBLP:conf/ccs/ZhuB21,DBLP:conf/msr/DingG21}, and the effectiveness and limitations of fuzzing techniques~\cite{DBLP:conf/icse/LiyanageBTL23,DBLP:conf/icse/BohmeSM22,DBLP:conf/icse/KloosterTBHB23,DBLP:conf/ccs/KleesRCW018}. While these studies provided valuable overall insights into fuzzing, continuous fuzzing, and fuzzing bugs, they have not considered the influence of programming languages, leaving language-specific characteristics  unexplored.

Zhu\etal\cite{DBLP:conf/ccs/ZhuB21} analyzed approximately 23,000 bugs reported in OSS-Fuzz and found that about 77\% of them were introduced by recent code changes. They observed that the proportion of these regression bugs increased in projects that had been fuzzed continuously over long periods. 
Ding\etal\cite{DBLP:conf/msr/DingG21} analyzed 23,907 bugs detected by OSS-Fuzz to investigate their characteristics (such as bug types, severity, and flakiness). They found that 52\% of the bugs primarily harmed system availability, such as crashes, timeouts, and out-of-memory errors. The median time from introduction to detection and from detection to fix is 5 days and 5.3 days, respectively. 
Keller\etal\cite{DBLP:conf/msr/KellerMM23} studied 44,102 OSS-Fuzz issues and their corresponding commits. They found that bug discovery rates decreased over time in most projects and that approximately 46\% of bugs were fixed by the same developers who introduced them.

Nourry\etal\cite{DBLP:journals/tosem/NourryKSSK25} investigated the issue of build failures in fuzzing and reported that approximately 9.4\% of OSS-Fuzz projects experience build failures without distinguishing between languages. 
Shirai\etal\cite{shirai2025tse} studied the evolution of fuzzing coverage, using 1 million fuzzing sessions from 808 projects on OSS-Fuzz. They showed that code coverage tends to increase continuously and that bug detection rates rise during coverage fluctuations.

\subsubsection{Language-Specific Analysis}
Several studies have conducted extensive analyses on various aspects of fuzzing itself (\eg coverage, efficacy, evolution) in a specific language or several related languages such as C and C++~\cite{DBLP:journals/corr/abs-2505-22052}.

Liyanage\etal\cite{DBLP:conf/icse/LiyanageBTL23} analyzed reachable coverage during long-term fuzzing using 32 real-world C programs. They observed that branch coverage increased linearly on a logarithmic time scale.
B{\"{o}}hme\etal\cite{DBLP:conf/icse/BohmeSM22} analyzed the reliability of coverage-based fuzzer benchmarking on a large variety of widely-used open-source C programs from different domains. They revealed that there is a strong correlation between the coverage achieved and the number of bugs found by a fuzzer. 
Nourry\etal\cite{DBLP:conf/icsm/NourryKAMK24} analyzed the adoption and evolution of fuzzing in 11,341 Go projects. Their results showed that the adoption of fuzzing remained limited, being used in only 3.15\% of all testing functions.

Hassler\etal\cite{DBLP:journals/corr/abs-2505-22052} analyzed over 100 vulnerabilities injected into C/C++ programs to evaluate 13 fuzzers. They revealed that fuzzing for C/C++ is most effective in detecting vulnerability types related to memory corruption and resource management errors, whereas it is less effective in identifying numerical computation errors.
Klooster\etal\cite{DBLP:conf/icse/KloosterTBHB23} studied the practical effectiveness of fuzzing and showed that even short fuzzing sessions of 10-30 minutes can achieve comparable detection effectiveness to long-running\blue ones\black.

\smallskip
Much of the prior work has also focused their analysis on the kinds of bugs that tend to get introduced into specific languages. These studies are first and foremost language-centric, meaning that they focus on the characteristics of the language unlike cross-language analysis studies which focus on bug characteristics.

For example, Imtiaz\etal\cite{DBLP:journals/corr/abs-2104-04385} analyzed vulnerabilities detected in C and C++ projects. Their results showed that approximately 40\% of the vulnerabilities were memory-related, and that the average CVSS3 score (\ie a measure of vulnerability severity) was 9.3 for memory-related vulnerabilities, which is higher than 6.6 for non-memory ones.
Zhang\etal\cite{DBLP:conf/qrs/ZhangFZDX24} analyzed 790 bugs from Rust projects and identified their root causes. They found that only 0.9\% of these bugs were security vulnerabilities, 53\% were caused by algorithm implementation errors, and only 3.8\% were related to memory issues.
Oh\etal\cite{DBLP:conf/sigsoft/OhO22} analyzed errors in real-world Python applications. Their results showed that type error exceptions accounted for about 30\% of all cases and were the most frequent, and that approximately 30\% of these required more than one month to fix, indicating a tendency toward prolonged resolution.

\subsubsection{Cross-Language Analysis}
While fuzzing research has yet to compare bugs across languages, many studies conduct bug-centric analyses to investigate how specific kinds of bugs manifest in different languages, how bug-fixes are implemented in different languages, or try to characterize vulnerabilities and bugs across multiple languages \cite{DBLP:conf/sigsoft/NikitopoulosDLM21, DBLP:journals/cacm/RayPDF17, DBLP:conf/wcre/KochharWL16, DBLP:conf/icse/NanzF15}.

Nikitopoulos\etal\cite{DBLP:conf/sigsoft/NikitopoulosDLM21} constructed a multilingual CVE dataset and organized the collected CVEs based on CWE categories. Their analysis showed that CWE-119 (\ie Bounds of a Memory Buffer) was the most common category in C language files, whereas CWE-79 (\ie Cross-site Scripting) was the most prevalent in PHP files.
Ray\etal\cite{DBLP:journals/cacm/RayPDF17} analyzed the number of defect-fixing commits across 17 programming languages and 729 projects. They showed that although the influence of programming languages on the number of defect fixes was minor, there was a strong correlation between languages and specific bug types, such as memory errors and concurrency issues.
Spinellis\etal\cite{DBLP:conf/oopsla/SpinellisKL12} investigated the behavior of fuzzing across programming languages in response to small changes in source code.
Their results show that in strongly typed systems, program changes can be detected at compile time, whereas in weakly typed systems, the program may still execute and produce incorrect outputs. 
Kochhar\etal\cite{DBLP:conf/wcre/KochharWL16} investigated the bug trends in projects developed using multiple programming languages. They found that using multiple languages increases both the number and diversity of bugs. 
Nanz and Furia~\cite{DBLP:conf/icse/NanzF15} compared solutions to 745 programming tasks from Rosetta Code, a multi-language code repository. They examined the fault-proneness of different languages and found that Go was the least fault-prone, whereas Python and Ruby tended to be the most fault-prone.

\subsubsection{Summing Up}
Related work indicates that fuzzing research primarily targets single-language environments (\eg C, C++, or Go). In contrast, vulnerability analyses frequently compare across multiple languages, establishing a correlation between language choice and distinct bug patterns. However, most large-scale empirical studies on continuous fuzzing have been language-agnostic. Given the fundamental differences in language design and safety guarantees—such as Rust emphasizing memory safety versus C/C++ providing low-level control—it remains uncertain whether general fuzzing effectiveness and findings apply across all languages. This ambiguity highlights the necessity for a comprehensive cross-language analysis of how fuzzing characteristics and bug detection efficiency are shaped by the programming language used.


\begin{figure*}[!t]
  \centering
    \includegraphics[width=0.65\textwidth]{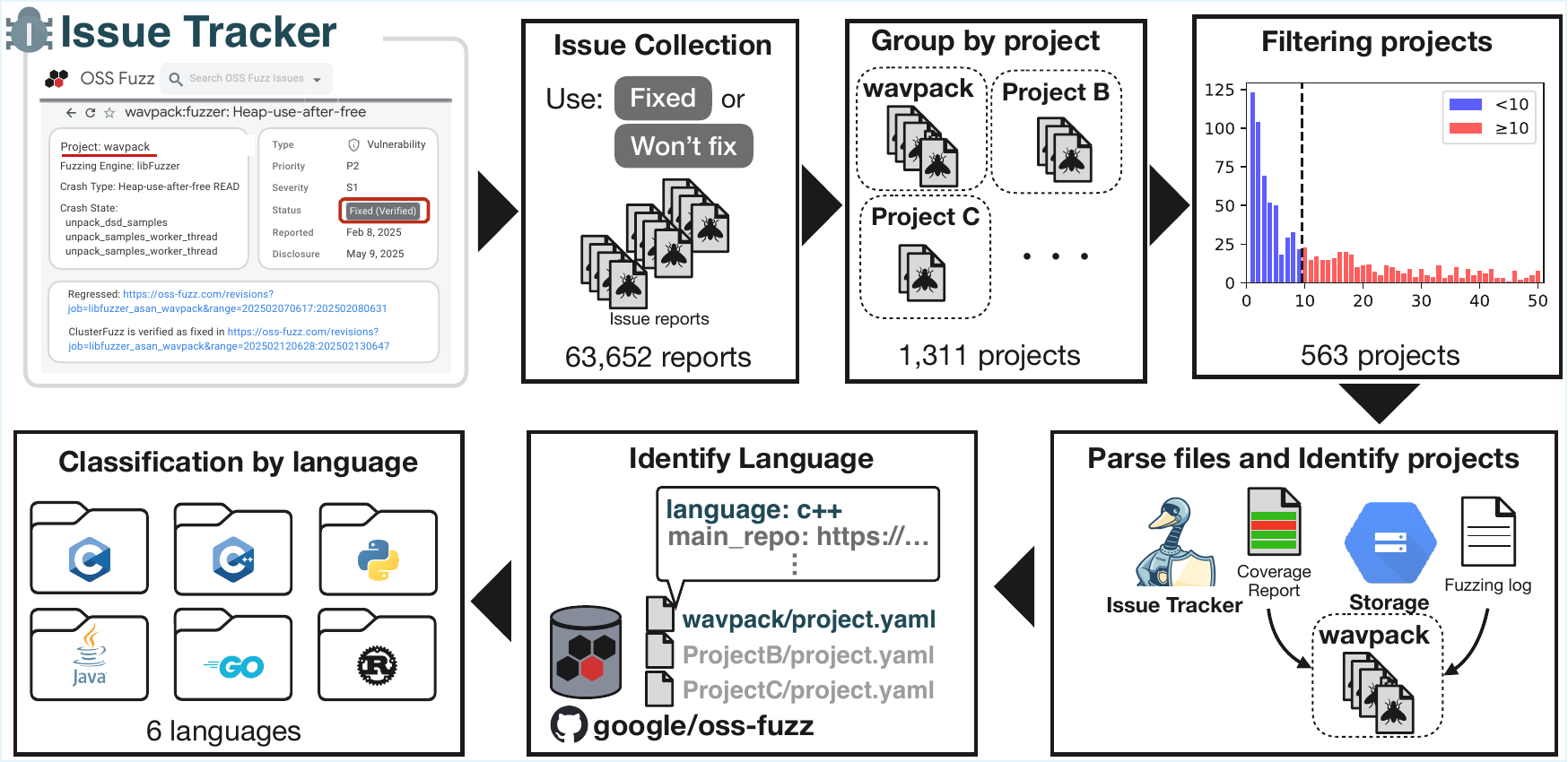}
    \caption{Overview of the data collection}
  \label{fig:overview}
\end{figure*}

\section{Data Collection}\label{sec:studydesign}
This section describes the procedures for data collection.
\autoref{fig:overview} depicts the overview of the data collection process.

\subsection{Issue Reports}\label{sec:project_collection}
This study examines fuzzing bugs (\ie bugs discovered during fuzzing sessions). We collect issue reports from OSS-Fuzz's issue tracker.\footnote{\url{https://issues.oss-fuzz.com/issues}}
Specifically, we collect all issue reports submitted by \ToIssueCollectDate, totaling \TotalIssueReport reports. We then filter for reports with status ``Fixed'' or ``Won't fix'' (\ie non-Open issues) to focus on conclusively resolved issues, yielding \ClosedIssues reports.

Next, we select appropriate projects to be analyzed. 
As of September 2025, \TotalFuzzingProject OSS projects are continuously fuzzed by OSS-Fuzz. However, some projects lacked sufficient issue reports for reasons such as having only recently begun fuzzing. Given the continuous operation of OSS-Fuzz, finding almost no bugs over an extended period is improbable and likely indicates misconfigurations or ineffective fuzz targets (e.g., shallow code coverage) rather than the absence of bugs. Because our analysis computes median values for projects grouped by programming language, we selected only projects with at least ten detected fuzzing bugs. This results in \TargetProject projects (\TargetProjectRate\% of the total).

\begin{table*}[t]
\small
  \centering
  \caption{Summary of studied projects and collected data}
  \label{tab:project_summary_by_language}
  \begin{tabular}{lrrrrrr}
    \toprule
    Language & \# Projects & \# Issues & \# Build Logs & \# Coverage Reports & Median Commits & Median LOC \\
    \midrule
    C           & 78  &  5,364 & 139,759 &  89,064 & 6,151 & 199,011 \\
    C++         & 264 & 47,128 & 591,651 & 451,767 & 4,578 & 126,216 \\
    Go          & 74  &  3,180 & 109,425 &  20,349 & 5,451 & 133,691 \\
    Java        & 67  &  2,339 &  66,608 &  47,101 & 3,239 &  77,558 \\
    Python      & 41  &    769 &  44,750 &  30,236 & 1,397 &  25,602 \\
    Rust        & 35  &  2,664 &  47,055 &  30,498 & 3,924 &  79,096 \\
    \midrule
    \textbf{Total} & \textbf{559} & \textbf{61,444} & \textbf{999,248} & \textbf{669,015} & -- & -- \\
    \bottomrule
  \end{tabular}
\end{table*}



\subsection{Fuzzing Build Logs and Coverage Reports}\label{sec:Fuzzing_session}
In RQ1, we use the number of sessions performed by OSS-Fuzz to calculate the average number of fuzzing bugs found by fuzzing sessions.
To measure it, we first collect all the OSS-Fuzz build logs (\AllBuildLog logs) for each project by scraping from Google Cloud Storage of OSS-Fuzz.\footnote{We needed to collect all because the information in each build log (\eg the build type and the target project) cannot be identified without parsing it.} We then parsed the log files, identified the projects, and filtered logs only for the studied repositories. In the end, we collected \FinalBuildLog build logs from the studied repositories.

Additionally, we use coverage reports in RQ4, which show the extent \blue to which \black the fuzzing sessions explore the code base. These coverage reports in OSS-Fuzz are generated daily (Note: not after every revision). We obtained \FinalCoverageReport coverage reports in total.

\subsection{Classification by Language}\label{sec:classification}

We identified the primary language used in each project and categorized the issues accordingly. We determined the main language using project descriptions from the OSS-Fuzz GitHub repository.\footnote{\url{https://github.com/google/oss-fuzz/blob/master/projects/abseil-cpp/project.yaml}} \red For the 10 projects where the configured language was changed, the change was likely a correction from the default setting to the project's main language or to add support for an additional language. \black After grouping by language, we found insufficient numbers of Swift and JavaScript projects (2 and 2 projects, respectively) and thus excluded them from our analysis. 
\autoref{tab:project_summary_by_language} presents summary statistics of the projects used in this study. 
It is worth noting that most of the issues were discovered by \texttt{libfuzzer}, which is the default fuzzer used by OSS-Fuzz. Specifically, \texttt{libfuzzer} detected 88.6\%-100\% of issues in C, Go, Java, Python, and Rust, versus 58.9\% in C++ (32\% of C++ reports do not mention which fuzzer was used).


\section{Research Questions}\label{sec:results}
\subsection*{\rqA}\label{sec:rqa}

\noindent\textbf{Motivation.}
Shirai\etal\cite{shirai2025tse} investigated the frequency of bugs discovered through continuous fuzzing, revealing that the bug detection rate fluctuates around 5\%. However, their study did not distinguish which languages were used in the studied projects. Different results may be obtained when examining modern software ecosystems that rely on diverse programming languages with varying memory management paradigms, type systems, and security features. To address this gap, we examine the frequency of fuzzing bugs in different programming languages.

\smallskip
\noindent\textbf{Approach.} 
We investigate how frequently fuzzing sessions find bugs. To measure the bug detection rate, we count the number of bugs found in OSS-Fuzz and divide by the total number of fuzzing builds. We repeat this calculation for each project, aggregate the results by programming language, and calculate the median value for each language. To check if different languages have significantly different detection rates, we apply the Kruskal–Wallis test ($\alpha=0.05$)~\cite{ac6c544c-0197-38bd-8c06-ec4f655ff4fd}. The Kruskal–Wallis test is a non-parametric statistical method that assesses whether samples originate from the same distribution, without assuming normality or equal variances. Since the Kruskal-Wallis test can only detect whether differences exist among the distributions, we apply the Mann-Whitney U test\cite{Mann1947OnATest} with Holm correction~\cite{29def780-e117-38f0-8afb-edf384af3fad} ($\alpha$=0.05) as a post-hoc test to identify which pairs of distributions differ significantly. Note that we apply this same statistical procedure (Kruskal-Wallis test followed by Mann-Whitney U test with Holm correction, $\alpha$=0.05) throughout RQ1-RQ4 whenever comparing distributions across different categories.

\begin{figure}[t]
\centering
\includegraphics[width=0.9\linewidth]{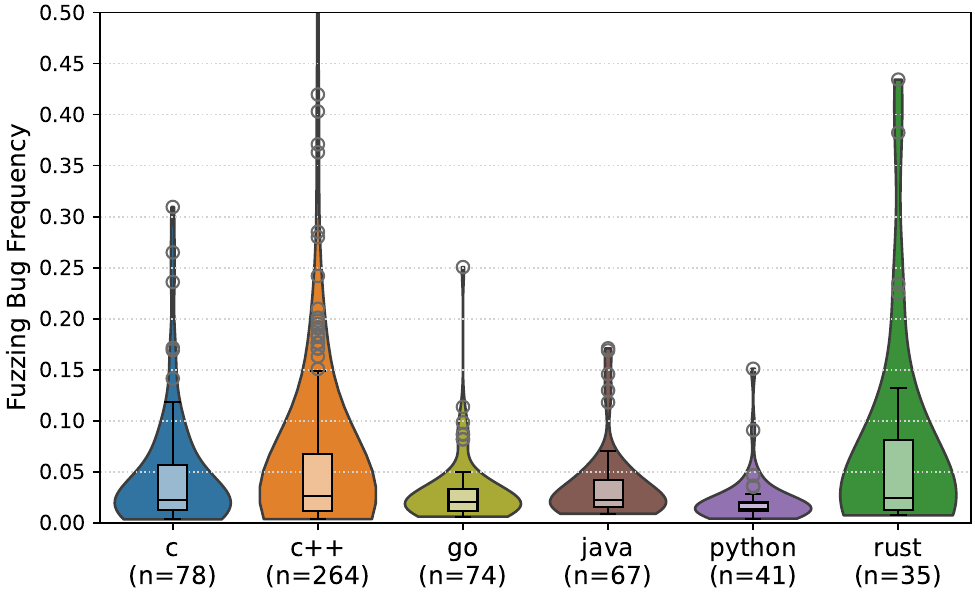}
\caption{Detection frequency of all fuzzing bugs relative to fuzzing runs.}
\label{fig:rq_two_frq_all}
\end{figure}

\smallskip
\noindent\textbf{Results. \textit{C, Go, and Java show similar median bug detection rates. C++ and Rust exhibit higher median bug detection with wider spread, whereas Python shows the lowest median and narrowest spread.}}
\autoref{fig:rq_two_frq_all} shows the distribution of total fuzzing bugs per build across six programming languages. We observe that the overall detection rates are low in all languages, with median bug detection rates around 0.02 fuzzing bug per build, as indicated by the horizontal lines in the box plots.

When examining the median values for each language, we observe subtle differences. First, the median values for C, Go, and Java are remarkably similar (approximately 0.02 fuzzing bugs per build), while C++ and Rust exhibit slightly higher medians (0.025), and Python show slightly lower medians (0.014). Despite similar median values, the languages differ significantly in their variability. C++ and Rust exhibit the widest distributions with long tails extending up to 0.15-0.45 fuzzing bug per build. C and Java show moderate variability, while Go displays a slightly more compact distribution. Python demonstrates the least variability with a narrow distribution centered near zero.

A Kruskal–Wallis test confirmed statistically significant differences across languages \blue ($H = 14.28$, $p = 0.014$), \black indicating that fuzzing bug detection rates vary by language. With the post-hoc tests to examine the difference between each pair, we observed statistically significant differences between \blue C++ and Python, and between Java and Python. \black

\begin{dbox}
\textbf{Answer to RQ1: }While C, C++, Go, and Rust show similar median fuzzing bug detection rates (0.02 fuzzing bugs per build), they differ substantially in variability, with C++ and Rust exhibiting the widest distributions and Python showing the most concentrated pattern with the lowest overall rate.
\end{dbox}




\subsection*{\rqB}\label{sec:rqb}
\smallskip\noindent\textbf{Motivation.}
Previous studies~\cite{DBLP:journals/corr/abs-2505-22052,DBLP:conf/msr/DingG21,DBLP:conf/msr/KellerMM23} have investigated the characteristics of fuzzing bugs. Hassler\etal\cite{DBLP:journals/corr/abs-2505-22052} compared fuzzers and static analysis tools, revealing fuzzers are particularly effective at detecting memory-related vulnerabilities such as bounds violations and use-after-free errors.
However, these findings may not apply to memory-safe languages like Rust, where compile-time safety guarantees prevent entire classes of those vulnerabilities. The fundamental differences in memory management between C and Rust likely result in distinct fuzzing bug patterns.
While C may reveal more memory corruption issues, Rust would be expected to detect logic errors or unsafe block vulnerabilities. Even if memory-related vulnerabilities are detected in Rust, the severity may be mitigated by its safety mechanisms.
Therefore, RQ2 examines OSS-Fuzz data across multiple languages to understand how language design influences vulnerability discovery rates.

\smallskip\noindent\textbf{Approach.} We measure three metrics: \rqaaOne, \rqbbTwo, and \rqbbOne, as follows.

\noindent\textit{$\mathrm{i}$) \rqaaOne } : 
Once a defect is detected by the fuzzers, OSS-Fuzz then classifies it into one of two types: ``Bug'' and ``Vulnerability''. ``Bug'' indicates an unexpected or incorrect program behavior and ``Vulnerability'' refers to a privacy or security issue as defined by Google’s security guidelines.\footnote{\url{https://developers.google.com/issue-tracker/concepts/issues\#type}} Using the \texttt{Type} field found in issue reports, we calculate the distribution of each type across all fuzzing bugs, grouped by language. 

\noindent\textit{$\mathrm{ii}$) \rqbbTwo } :
OSS-Fuzz automatically\footnote{\url{https://github.com/google/clusterfuzz/blob/5b7f6aa8daf56b3f33508855f17fe77570baf19a/src/clusterfuzz/_internal/crash_analysis/severity_analyzer.py}} assigns severity levels to crashes when creating issue reports. Each issue report contains a \texttt{Severity} field that ranges from ``S1'' (most severe) to ``S4'' (least severe). We measure the distribution of severity levels for each language group.
It is worth noting that most of the issues with ``Bug'' tag have ``S4'' level so we only study the severity of issues with the ``Vulnerability'' tag.

\smallskip\noindent\textit{$\mathrm{iii}$) \rqbbOne } : 
Fuzzing bugs can be classified into multiple types based on their CWE-ID~\cite{CWE-Official-Web,DBLP:conf/msr/DingG21}. Issue reports in OSS-Fuzz contain a \texttt{Crash Type} field that reveals how crashes manifest. We classify them, employing the highest-level categories (\ie Pillar categories) from the CWE VIEW: Research Concepts.\footnote{\url{https://cwe.mitre.org/data/definitions/1000.html}} \autoref{tab:cwe_id} lists the CWE-IDs and their corresponding names used in this study. We calculate the distribution of CWE categories across all fuzzing bugs grouped by programming languages.

\begin{table*}[th]
\small	
  \centering
  \begin{threeparttable}
    \caption{Types of CWE }
    \begin{tabularx}{\linewidth}{l l X X}
      \toprule
      \textbf{Label}& \textbf{CWE-ID} & \textbf{Description} & \textbf{Example Crash Type}\\
      \midrule
      Resource Management & CWE-664 & Improper Control of a Resource Through its Lifetime 
     & Timeout, Out-of-memory, buffer-overflow  \\ 
    Incorrect Calculation & CWE-682 & Incorrect Calculation
      & integer-overflow, divide-by-zero\\ 
      Control Flow & CWE-691 & Insufficient Control Flow Management
      & assert, stack-overflow, ill\\ 
      Protection Mechanism & CWE-693 & Protection Mechanism Failure
      & security-exception, \\ 
      Exceptinal Handling & CWE-703 & Improper Check or Handling of Exceptional Conditions
      & abrt, unexpected-exit, fatal-signal\\ 
      Neutralization & CWE-707 & Improper Neutralization
      & index-out-of-range, slice-bounds-out-of-range\\ 
      Coding Standards & CWE-710 & Improper Adherence to Coding Standards
      & null-dereference, undefined-shift\\
      \bottomrule
    \end{tabularx}
    \begin{tablenotes}[flushleft]
        \footnotesize
        \item \red\textit{Note:} We distinguish stack-overflow from buffer-overflow based on their root causes (\ie control recursion vs. memory bounds).\black
    \end{tablenotes}
    \label{tab:cwe_id}
  \end{threeparttable}
\end{table*}

\begin{figure}[t]
\centering
\includegraphics[width=0.8\linewidth]{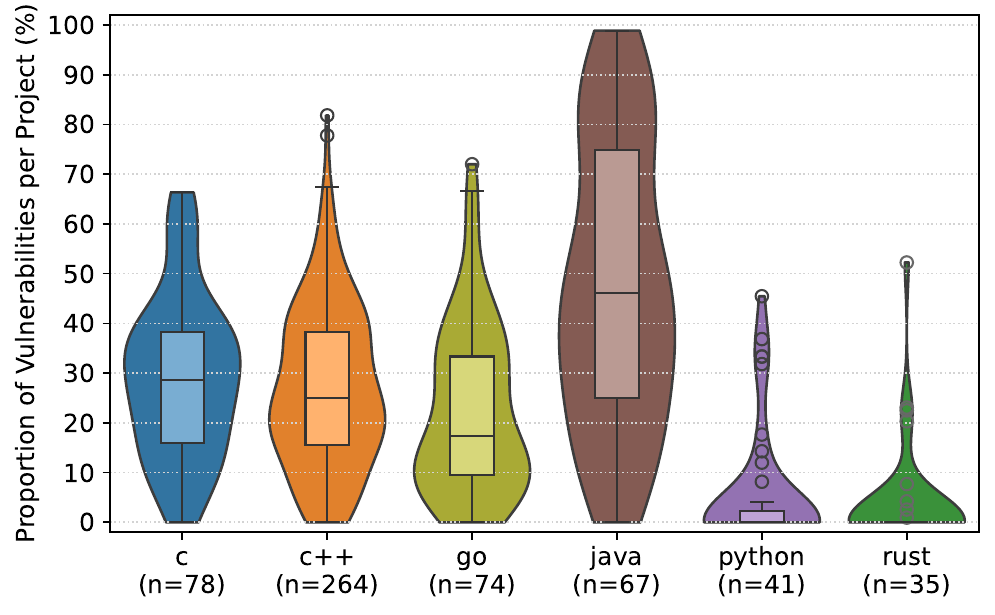}
\caption{Proportion of fuzzing bugs with Vulnerability label.}
\label{fig:rq_one_vul_rate}
\end{figure}

\smallskip\noindent\textbf{Results. \textit{Vulnerability rates are lowest in Python and Rust, higher in C and C++, while Java has the highest rate overall. }}
\autoref{fig:rq_one_vul_rate} shows the vulnerability ratio among fuzzing bugs across different languages. With an overall vulnerability rate of \rqOneVulAllRate\%, C and C++ exhibit ratios close to the overall rate. These languages demonstrate not only high occurences of fuzzing bugs (as shown in RQ1) but also a substantial number of vulnerabilities. Notably, Java projects have the highest vulnerability ratio, with \blue nearly half (46.3\%) \black of their fuzzing bugs classified as vulnerabilities. In contrast, Python and Rust have a significantly lower proportion of vulnerabilities. Notably, while RQ1 revealed that Rust projects detect numerous fuzzing bugs, their vulnerability ratio remains below 10\%. We observed statistically significant differences detected by Kruskal-Wallis tests and also between each pair, except C and C++ as well as Rust and Python.

\begin{figure}[t]
\centering
\includegraphics[width=0.8\linewidth]{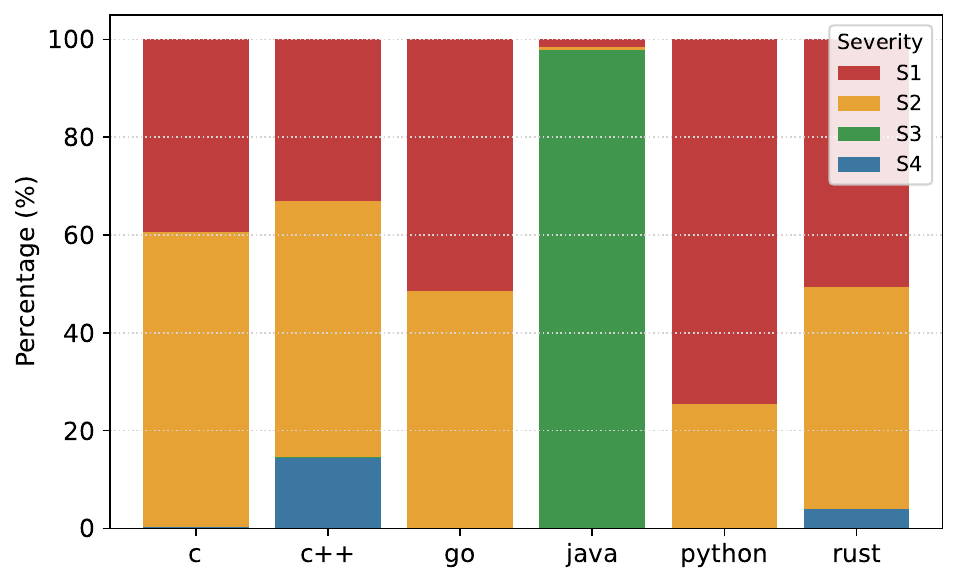}
\caption{Severity (S1–S4) distribution by language for Vulnerability-labeled fuzzing bugs.}
\label{fig:rq_one_severity_vuln}
\end{figure}

\smallskip\noindent\textbf{\textit{Severity varies dramatically by language, with Java showing almost exclusively medium-severity vulnerabilities (S3) while all other languages are dominated by critical and high-severity \red vulnerabilities \black (S1-S2).}}
\autoref{fig:rq_one_severity_vuln} shows the severity distribution of vulnerability-labeled fuzzing bugs across six programming languages. We observe several different tendencies by languages. The most striking pattern emerges with Java, which exhibits an entirely unique profile where nearly 100\% of its vulnerability-labeled fuzzing bugs fall into the S3 (medium severity) category. This unique pattern may be attributed to Java's memory-managed nature and built-in safety features that prevent many of the critical vulnerabilities found in other languages~\cite{secure-coding-java-se}. 

In contrast to Java, most other languages are dominated by high-severity vulnerabilities. C, C++, Go, and Rust show predominantly S1 (highest severity) and S2 vulnerabilities, with S1 typically accounting for 30-50\% of the vulnerabilities discovered through fuzzing. Python also exhibits an exceptionally high proportion of S1 vulnerabilities at approximately 75\%, markedly exceeding the rates observed in other languages.

The results also reveal that S4 (lowest severity) vulnerabilities are quite rare across all languages. Only C, C++ and Rust show any S4 vulnerabilities at all, and even in these cases, they represent a very small percentage, less than 20\% of the total.

\begin{figure}[t]
\centering
\includegraphics[width=0.8\linewidth]{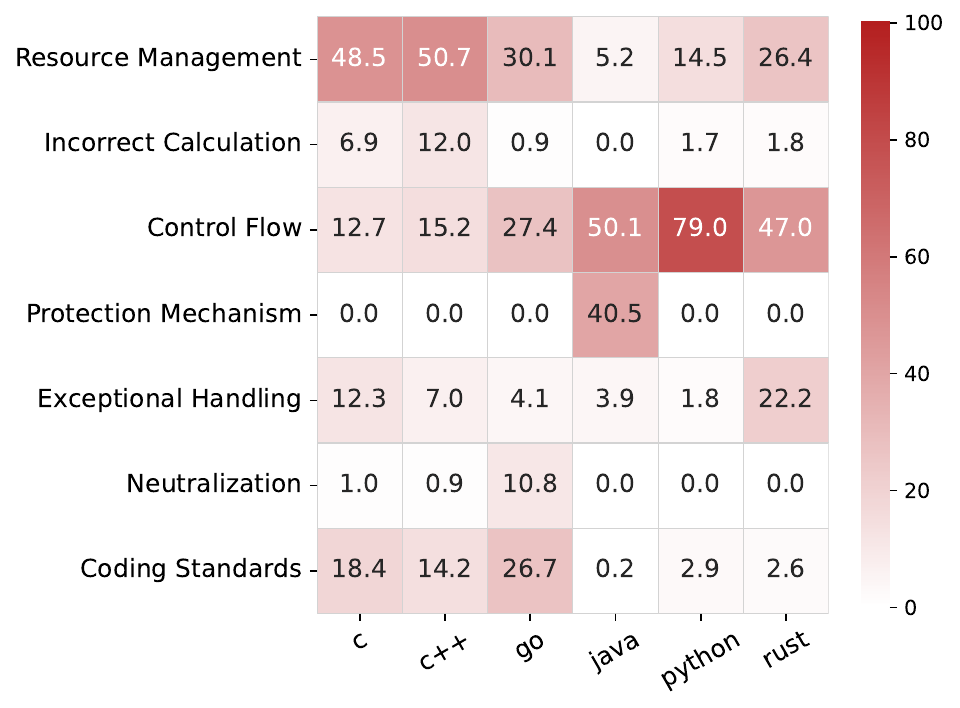}
\caption{Proportion of all fuzzing bugs by crash types across languages.}
\label{fig:rq_one_cra_type_all}
\end{figure}

\smallskip\noindent\textbf{\textit{Memory-unsafe languages (C, C++) are dominated by Resource Management bugs, whereas managed languages (Java, Python) primarily suffer from Control Flow–related bugs.}}
\autoref{fig:rq_one_cra_type_all} shows the distribution of crash types across programming languages as a heatmap, with darker red indicating higher values. 

Memory-unsafe languages exhibit distinct patterns. C and C++ show the highest Resource Management bug rates at 48.6\% and 50.7\%, respectively, reflecting their manual memory management requirements.
In contrast, Java, Python, and Rust are predominantly affected by Control Flow bugs. This trend is most pronounced in Python, where Control Flow bugs account for 79.0\% of all fuzzing bug. Java also shows 50.2\% Control Flow bugs and is notably the only language exhibiting Protection Mechanism bugs. 
Rust demonstrates interesting characteristics. Although Control Flow bugs are dominant at 47.0\%, Resource Management bugs also account for 26.4\%.
Since this is significantly lower than in memory-unsafe languages, Rust's ownership system helps prevent resource-related crashes but does not fully eliminate logic-level vulnerabilities. Go, on the other hand, shows a more balanced distribution of bug types, without strong concentration in any single category.











\begin{dbox}
\textbf{Answer to RQ2: }
Java projects have the most vulnerabilities but mainly medium-severity ones, while Python and Rust have fewer but more critical issues. C, C++, and Go show similar patterns. For crash types, C, C++, and Go mainly have Resource Management bugs, while Java, Python, and Rust have Control Flow bugs.
\end{dbox}

\subsection*{\rqC}\label{sec:rqc}
\smallskip\noindent\textbf{Motivation.}
Reproducing bugs consistently is important for developers to fix them. To facilitate bug fixing, fuzzing infrastructure like ClusterFuzz provides the necessary information to reproduce issues and confirm that fixes are correct. However, reproducibility is not always guaranteed~\cite{clusterfuzz_fixing_bug}. Ding\etal\cite{DBLP:conf/msr/DingG21} showed that 13\% of reported issues in OSS-Fuzz are unreproducible, which can hinder the bug-fixing process and waste developer effort. While their study provides valuable insights into the overall reproducibility problem, they did not distinguish between different programming languages. Given that different languages have varying memory management models, runtime behaviors, and debugging toolchains, reproducibility rates may differ significantly across languages.

\smallskip\noindent\textbf{Approach.} 
OSS-Fuzz classifies each crash as either ``Reproducible'' or ``Unreproducible'' in its reports. Unreproducible crashes occur intermittently due to factors like race conditions or uninitialized memory. We calculate the proportion of ``Reproducible'' bugs for each studied language. 

\begin{figure}[t]
\centering
\includegraphics[width=0.8\linewidth]{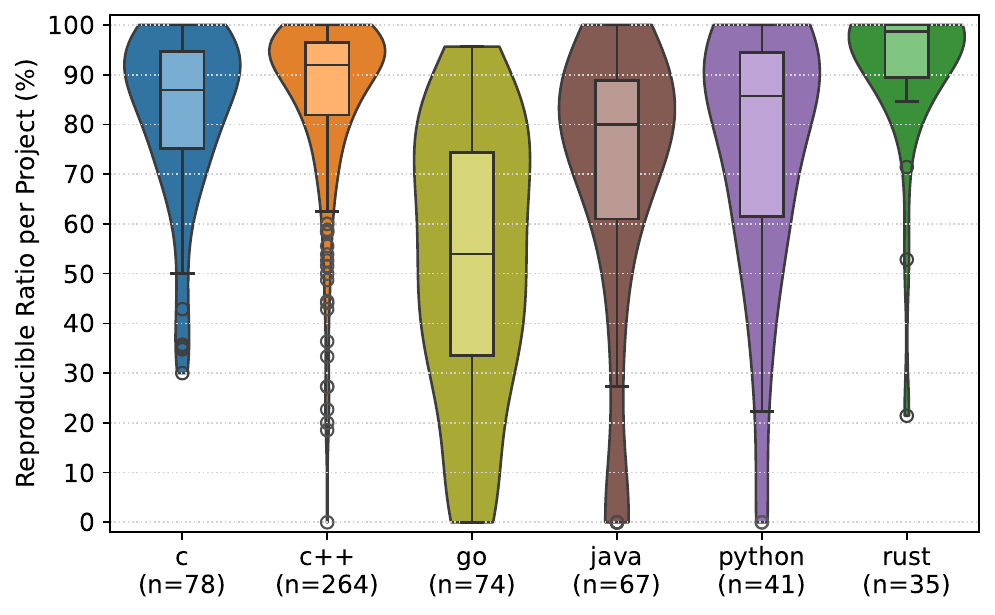}
\caption{Reproducibility rate of crashes across languages.}
\label{fig:rq_one_flaky_bug}
\end{figure}

\smallskip\noindent\textbf{Results. \textit{Reproducibility varies across programming languages, with Go showing the lowest reproducibility rate (median about 50\%), and C/C++/Python/Java performing better (80-95\%).}}
\autoref{fig:rq_one_flaky_bug} displays the distribution of reproducibility ratio  per language using violin plots with embedded box plots. 
Looking at the overall results, there are significant variations in reproducibility rates across the different programming languages. The majority of projects have relatively high reproducibility ratios, but some outliers reach very low percentages. Go shows the lowest reproducibility, with a median reproducibility ratio of about 50\% and a broad distribution that extends across nearly the entire range. C, Java, and Python exhibit similar patterns with medians around 85\% of reproducible crashes and fairly concentrated distributions, suggesting comparable reproducibility challenges. C++ presents an interesting case with a high median around 90\% but notable outliers reaching as low as 10\%, indicating that while most C++ projects are reproducible, some face significant challenges. 

\smallskip\noindent\textbf{\textit{Rust projects demonstrate the best reproducibility characteristics with the highest \blue median \black around 98.8\% and a narrow distribution, indicating that almost all fuzzing bugs are consistently reproducible.}}
Among the languages studied, Rust projects demonstrate exceptional fuzzing bug reproducibility with the highest observed reproducibility ratio (median 98.8\%) and the most concentrated distribution near 100\% among all studied languages. This superior performance suggests that Rust's design features and toolchain effectively promote deterministic and consistent behavior compared to other programming languages. 
We observed statistically significant differences detected by the Kruskal-Wallis test. 
Specifically in post-hoc tests, we found significant differences (p < 0.05) 
between
\blue all language pairs, except for C vs. C++, C vs. Python, and Java vs. Python. \black

\begin{dbox}
\textbf{Answer to RQ3: }
Rust projects demonstrate the best reproducibility with a median ratio of reproducible crashes around 98.8\% and a tight distribution near 100\%, while Go projects perform worst with a median ratio near 50\% and a broad distribution.
\end{dbox}


\subsection*{\rqD}\label{sec:rqd}
\smallskip\noindent\textbf{Motivation.} 
Coverage-Guided Fuzzing (CGF) is the primary fuzzing strategy adopted by OSS-Fuzz, leveraging knowledge of the codebase and coverage information to explore a broader range of code paths. Higher code coverage is generally regarded as an indicator of fuzzing effectiveness for a given software system. Shirai\etal\cite{shirai2025tse} reported that improving coverage is more likely to help detect bugs, demonstrating a direct relationship between coverage and bug discovery. Another critical metric for evaluating fuzzing efficiency is Time-to-Detect, which refers to the \blue time \black elapsed between a bug's introduction and detection. Ding\etal\cite{DBLP:conf/msr/DingG21} showed that OSS-Fuzz detects the majority of identified regressions within a week, highlighting the practical value of rapid bug detection. However, it remains unclear whether fuzzing efficiency varies across programming languages.

\smallskip\noindent\textbf{Approach.}
We measure two proxy metrics for fuzzing efficiency as follows.

\smallskip\textit{$\mathrm{i}$ ) \rqccOne } : 
Patch coverage measures how much of the modified code in commits is exercised by test suites~\cite{DBLP:journals/pacmse/YildiranOLG24}. Fuzzers for continuous fuzzing are often evaluated based on how effectively they reach the newly modified code~\cite{DBLP:conf/asiaccs/CanakciMGJE22, DBLP:conf/uss/Xiang00JXLXW24, DBLP:journals/iet-sen/ZhangCCYZL23}.
To calculate this, we first identified the revisions and dates on which each target project was changed, based on the daily coverage reports. We then determined the modified code (the diff) for each of those revisions. Finally, we matched the coverage reports against the diff‐covered code regions in order to compute patch coverage.
We then calculate patch coverage by matching the modified code regions with the corresponding coverage reports.
To provide representative median values, we only selected projects for which patch coverage could be computed for at least 10 revisions in each project (\PatchCoverageProject projects).
Finally, we iteratively apply this process for each project, compute the median patch coverage, and aggregate the results by programming language to analyze patch coverage distributions. 

\smallskip\textit{$\mathrm{ii}$ ) \rqccTwo } : 
Time-to-detection (TTD) is the number of days between when fuzzing bugs are introduced and when they are detected by fuzzing. Similar to prior work by Ding\etal\cite{DBLP:conf/msr/DingG21}, we identified the introduction date of fuzzing bugs by checking the ``Regressed Revision'' field in issue reports, which indicates the build in which the bug first appeared. We marked the detection date as when the issue report was filed, then calculated the TTD as the days between these two dates. We compute the median TTD for each project and examine the distributions of these medians for each programming language.

Note, since this analysis relies on the availability of the ``Regressed Revision'' field of the issue report, we excluded issues where this information could not be extracted. We further filtered the projects that have fewer than 10 issue reports to obtain representative median values. As a result, this analysis targeted \RegressedProjects projects and \RegressedIssues issues in total. It is worth noting that all reports with the ``vulnerability'' tag in Go projects lacked the ``Regressed Revision'' entry. To keep our data consistent across languages, we therefore had no choice but to exclude GO projects from this analysis.

\begin{figure}[t]
    \centering
    \includegraphics[width=0.9\linewidth]{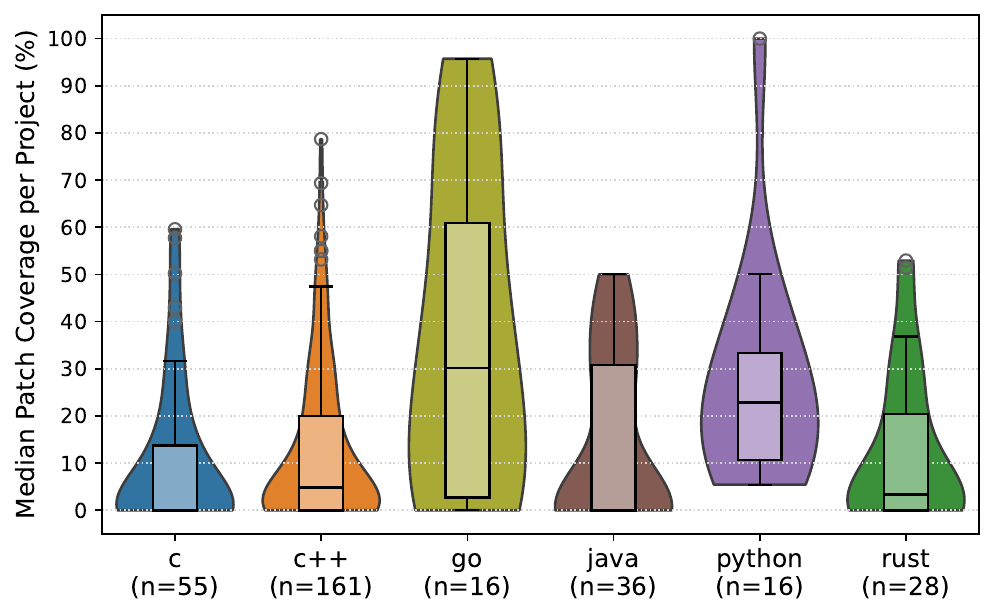}
    \caption{Distributions of Patch Coverage across languages.}
    \label{fig:rq_four_patch_coverage}
\end{figure}
\smallskip\noindent\textbf{Results. \textit{Patch coverage varies significantly across languages. Go and python show the highest coverage, with a median hovering around 30.2\% and 22.9\% respectively. In contrast, C, C++, Java, and Rust all have medians below 10\%, though some projects exceed 50\%.}}
\autoref{fig:rq_four_patch_coverage} shows the distribution of per-project patch coverage for each programming language. 
The results indicate a clear distinction between two groups: languages that maintain relatively high patch coverage (Go and Python) and those that exhibit consistently low coverage (C, C++, Java, and Rust). 
Notably, Go shows a relatively higher median patch coverage around 30.2\%, likely due to its built-in fuzzing support and simple language design. Similarly, Python achieves a comparatively high median of around 22.9\%. As the only dynamically typed language in our study, Python's runtime flexibility may allow fuzzers to explore diverse code paths more easily, though further investigation is needed to confirm this hypothesis. \blue While the Kruskal-Wallis test detects overall differences among distributions, post-hoc tests identify significant differences between Python and both C and C++, as well as between C and Go. \black

In contrast, several programming languages (C, C++, and Rust) show wide distributions, but their median patch coverage remains below 10\%.
This may be because of the high structural complexity of these languages which hinders fuzzers' ability to achieve high coverage. This could also be due to the high amount of fuzzing bugs (as shown by RQ1) in these languages preventing subsequent code execution (\ie the code execution is terminated by crashes).





\begin{figure}[t]
    \centering
    \includegraphics[width=0.9\linewidth]{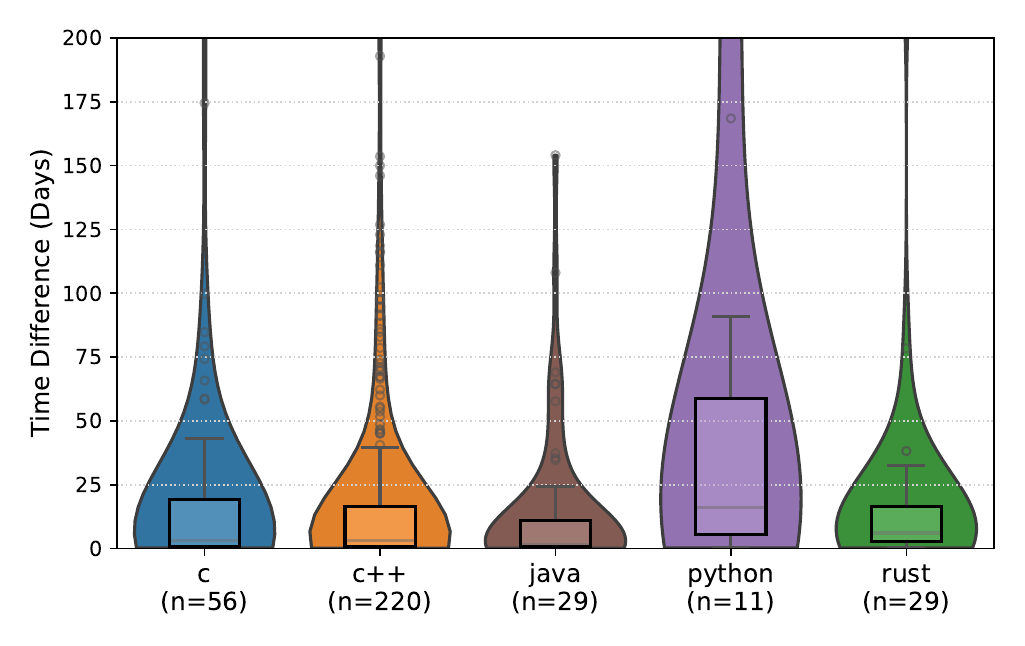}
    \caption{Distributions in the median time to detect fuzzing bugs discovered by each project, grouped by language.}
    \label{fig:rq_two_frq_ttd_proj}
\end{figure}

\smallskip\noindent\textbf{\textit{Time-to-detection (TTD) varies substantially across programming languages. C and C++ tend to detect bugs within about a week, whereas Java and Rust typically detect them within around three days. In contrast, Python shows a tendency toward much longer detection times.}}
\autoref{fig:rq_two_frq_ttd_proj} presents the distributions of median TDD time to detect fuzzing bugs across projects, grouped by programming language.

The C and C++ programming languages exhibit similar distributions and median values (7.0 and 9.5 days, respectively), indicating that bugs in these languages are typically detected by fuzzing within roughly one week. As shown in RQ2, these languages often involve memory-related issues that require complex inputs, which might contribute to longer detection times.

In contrast, Java and Rust show much shorter median TTDs of 2.8 and 3.9 days, respectively, indicating that bugs are detected more quickly. These languages tend to have a higher proportion of control-flow-related bugs in common. Similarly, Python projects also exhibit many control-flow-related bugs and higher patch coverage. Yet Python's median TTD is 26.8 days, showing that the detection process takes considerably longer than in other languages. This suggests that language-specific factors might influence detection speed, and future studies should investigate these differences.

A Kruskal–Wallis test confirmed that these differences across languages are statistically significant ($H = 183.3$, $p < 0.01$), indicating that the time-to-detection of fuzzing bugs varies significantly by programming language. The post-hoc tests detected significant differences \blue between all language pairs, except for Rust vs. Java. \black




\begin{dbox}
\textbf{Answer to RQ4: }
Fuzzing efficiency differs by programming language. Languages with higher patch coverage, like Go and Python, show longer detection times, while C, C++, Java, and Rust have lower coverage but faster detection. Though testing new code should enable early bug detection, this relationship does not hold consistently for fuzzing across languages.
\end{dbox}

\section{Discussion}
\subsection{Lessons Learned}

The central finding of this comprehensive empirical study is the conclusive confirmation that projects written in different programming languages have different characteristics and handling of fuzzing bugs. This section summarizes the resulting language-specific lessons learned, highlighting the distinct patterns for each language.

\texttt{C and C++} projects show moderate to slightly higher median bug detection rates (0.02-0.025 issues per build), with C++ exhibiting the highest variability. Crash types are heavily dominated by Resource Management bugs (48.6\% for C, 50.7\% for C++), reflecting manual memory handling requirements. Both languages maintain low median patch coverage ($\sim$10\%) but achieve quick detection times, with median TTD of 7.0 days for C and 9.5 days for C++.

\texttt{Go} projects face the greatest reproducibility challenge, with a median of only 50\%. However, Go achieves the highest median patch coverage at around 30.2\%. 

\texttt{Java} projects present a unique security profile: the highest vulnerability ratio, yet nearly 100\% are medium-severity (S3) due to memory management. Java achieves the fastest detection time with a median TTD of 2.8 days.

\texttt{Python} projects exhibit the lowest median bug detection rate (0.014) and its narrowest distribution. Control Flow bugs dominate at 79.0\%. Despite higher patch coverage (median around 22.9\%), Python has the longest median TTD at 26.8 days, indicating that coverage does not guarantee rapid detection.

\texttt{Rust} projects have a relatively high fuzzing bug detection rate but a significantly low vulnerability ratio (below 10\%). Rust demonstrates superior reproducibility with the highest median reproducible ratio (around 98.8\%).
It achieves quick detection with a median TTD of 3.9 days, similar to Java.

\smallskip
\textit{Consideration of Confounding Factors.} 
While our results exhibit clear differences across programming languages, the differences noted might not solely be attributable to language-specific problems. We utilized median values to help mitigate statistical bias introduced by individual project outliers and confounding factors. However, other factors such as project characteristics, might also impact our results and our observations may not stem exclusively from the programming language itself. It is important to acknowledge that all projects studied are well-established and selected by OSS-Fuzz. For instance, the Python projects analyzed are large with a median number of lines of code of 354,370 (See \autoref{tab:project_summary_by_language}). Future research should identify the biggest contributing factors underlying these observations, using established techniques such as Regression Discontinuity Design (RDD)~\cite{rrdAnalysis, rrdGuide, DBLP:conf/icsm/WesselSWSG20}.

\subsection{Implications}\label{sec:implications}

\subsubsection{Implications for Researchers}
Prior fuzzing research has predominantly focused on C or C++ projects, overlooking the fundamental impact of language differences. Since our comprehensive cross-language analysis demonstrates that programming languages fundamentally shape bug characteristics and detection efficiency, future studies must focus on language-specific bugs. This approach is necessary because different languages exhibit varying degrees of vulnerability resilience and severity, making it dangerous to generalize findings across languages. For instance, our results (RQ2) showed that C and C++ primarily suffer from Resource Management bugs, whereas Python and Rust tend to exhibit Control Flow bugs.
Researchers should also pay attention to the impact of low bug detection rates. For example, if we look at Python projects, our results showed that Python projects tend to have a low detection rate (median value of 0.014 issues per build) but also that a significant proportion of fuzzing bugs (79.0\%) found in these projects were Control flow Bugs. \textit{Researchers should therefore be careful about whether the high percentage of Control Flow bugs reflects a genuinely high bug rate for that specific type bugs, or whether it indicates that current fuzzing techniques are ineffective at detecting other types of defects in Python.}

\subsubsection{Implications for Fuzzing Platform Providers}
Our findings highlight the need for language-specific fuzzers, given that fuzzing sessions and detected bugs manifest differently across programming languages. 
\textit{Fuzzing platform providers should therefore offer specialized fuzzers for each supported language.} For example, fuzzers targeting C and C++ should prioritize detecting Resource Management bugs, which constitute nearly half of their vulnerabilities (48.6\% and 50.7\%, respectively). Conversely, fuzzers for interpreted languages like Python should be optimized to find Control Flow issues. Several studies have developed fuzzers for Rust~\cite{DBLP:journals/corr/abs-2505-02464,DBLP:conf/kbse/Jiang0Z21}, Python~\cite{DBLP:conf/ccs/LiYL0C23,DBLP:journals/corr/abs-2403-12723}, and Java~\cite{DBLP:conf/pldi/ChenSSSZ16,DBLP:conf/issta/PadhyeLS19}, which serve as viable alternatives.

Furthermore, given that patch coverage remains below 50\% across most languages (RQ4), especially in C, C++, Java, and Rust (which all have medians below 10\%), traditional fuzzing approaches using general Greybox Fuzzing Tools face significant limitations. \textit{Fuzzing providers should offer Directed Greybox Fuzzing (DGF)\cite{DBLP:conf/ccs/BohmePNR17} as an option.} DGF can focus on specific areas of code and can be directed to the changed lines of code in commits. The low coverage metrics indicate that general coverage-guided fuzzing (CGF) may fail to adequately test newly modified code, necessitating the use of directed approaches.

\subsubsection{Implications for Practitioners}
As observed, Java projects present a unique security profile: they have the highest overall vulnerability ratio among the languages studied, yet nearly 100\% of these detected issues are classified as S3 (medium severity). This pattern suggests that Java's built-in safety features, such as garbage collection and memory management, effectively prevent the most critical vulnerabilities common in other environments. Conversely, practitioners working with systems and scripting languages must prepare for inherently more severe vulnerabilities. C, C++, Go, and Rust primarily yield S1 (critical) and S2 (high severity) vulnerabilities. Python is particularly noteworthy, showing an exceptionally high proportion of S1 vulnerabilities, representing approximately 75\% of its detected defects. These differences extend to crash types as well, where C and C++ primarily suffer from Resource Management bugs, while Java, Python, and Rust tend to show more Control Flow bugs. 

Additionally, the language choice may dictate the efficiency of the bug remediation process, which is heavily influenced by reproducibility. Reproducing bugs consistently is essential for developers to successfully fix issues and confirm that the fixes are correct; conversely, unreproducible bugs waste development time and are less likely to be addressed. Our results revealed significant, language-dependent variations in the ratio of defects that were reproducible. Therefore, \textit{practitioners should choose languages depending on their project's security requirements, tolerance for critical vulnerabilities, and capacity for debugging reproducibility challenges.}
\section{Threats to Validity}\label{sec:threats_to_validity}

\smallskip\noindent\textbf{Threats to internal validity:}
Internal validity refers to the degree to which causal relationships can be established between variables without interference from confounding factors.
In our study, internal validity is threatened by selection bias in the data collection process.
We selected only projects where at least 10 fuzzing bugs were detected, which may introduce bias, as variations in detection rates could be attributed to project characteristics (e.g., projects with a tendency to frequently reveal bugs) rather than the factors we aim to investigate.

\smallskip\noindent\textbf{Threats to construct validity:}
Construct validity concerns whether the operational measures used in the study accurately represent the theoretical concepts being investigated.
In our study, construct validity is threatened by potential confounding factors, as project maturity or configuration could influence bug detection rates rather than the inherent properties of the programming language.
To mitigate this threat, we calculated and compared the median values for metrics across projects, minimizing the statistical bias introduced by individual project outliers.

\smallskip\noindent\textbf{Threats to external validity:}
External validity addresses the extent to which the study's findings can be generalized to other settings, populations, or conditions beyond the specific context of the research.
In our study, external validity is threatened by the limited scope of our study population.
This study is limited to projects that are continuously fuzzed by OSS-Fuzz, a service provided by Google.
While OSS-Fuzz is the most widely used continuous fuzzing service today, it remains unclear whether our results can be generalized to other fuzzing tools, projects not participating in OSS-Fuzz, or commercial software development that does not incorporate continuous fuzzing into their CI/CD pipelines.
\red Lastly, we applied strict filtering to obtain reliable median values. Although this smaller sample size may limit generalizability, it protects the robustness of our results against statistical noise. \black

\section{Conclusion}\label{sec:conclusion}
This empirical study performed a comprehensive cross-language analysis to address the critical gap in previous language-agnostic fuzzing research, investigating how bug characteristics and detection efficiency vary across programming languages, utilizing \FinalIssue fuzzing bugs from \FinalProject OSS-Fuzz projects. 

Our findings confirm that programming language significantly influences the characteristics and handling of detected bugs, supporting the necessity of language-aware security analysis. While the median bug detection frequency (RQ1) and the fuzzing efficiency (Time-to-Detection, RQ4) remained generally consistent across languages—with median detection rates around 0.02 issues per build and TTDs between 5 and 25 days—the nature of the vulnerabilities differed sharply. Specifically, Java projects exhibited a distinct profile dominated by medium-severity (S3) vulnerabilities, contrasting with \red the predominance of critical and high-severity (S1/S2) vulnerabilities \black in C, C++, Go, Python, and Rust. 

Furthermore, language design impacted reproducibility (RQ3): Rust projects demonstrated superior build reproducibility with the high median reproducible ratio (around 95\%), while Go showed the poorest performance with a median near 50\%. These results demonstrate that language-specific safety features correlate with distinct security outcomes in the OSS ecosystem. 

Our future work should identify the biggest contributing factors underlying these observations, potentially utilizing established techniques such as Regression Discontinuity Design (RDD). We must also focus on language-specific bugs, investigating, for instance, whether the high percentage of Control Flow bugs in Python reflects a genuinely high bug rate for that specific type, or if it indicates that current fuzzing techniques are ineffective at detecting other types of defects in Python.

\begin{acks}
We gratefully acknowledge the financial support of JSPS for the KAKENHI grants (JP24K02921, JP24K02923, JP25K21359), as well as JST for the PRESTO grant (JPMJPR22P3), the ASPIRE grant (JPMJAP2415), and the AIP Accelerated Program (JPMJCR25U7).

\end{acks}

\balance
\bibliographystyle{ACM-Reference-Format}
\bibliography{references}


\end{document}